\documentclass{article}
\usepackage{spconf}
\usepackage{subcaption}
\usepackage{amsthm}
\usepackage{amsfonts}
\usepackage{enumitem}
\usepackage{amssymb}
\usepackage{xcolor}
\usepackage{url}
\usepackage{graphicx}
\usepackage{subcaption}
\usepackage[ruled,vlined]{algorithm2e}
\usepackage{xcolor}
\usepackage[normalem]{ulem}
\usepackage{amsfonts}
\usepackage{amsmath}
\usepackage{amssymb}
\usepackage{xcolor}
\usepackage{amsthm}
\usepackage{caption}

\newcommand{\R}{\mathbb{R} }

\newcommand{\C}{\mathbb{C}}

\newcommand{\tr}{\text{trace}}

\renewenvironment{thebibliography}[1]{%
	\begin{oldthebibliography}{#1}%
		\setlength{\itemsep}{-0.75ex}%
	}%
	{%
	\end{oldthebibliography}%
}

\usepackage{setspace}
\begin{document}
	\title{ SPARSE MULTI-REFERENCE ALIGNMENT: SAMPLE COMPLEXITY AND COMPUTATIONAL HARDNESS \vspace{-.5em}} 
	%
	\name{Tamir Bendory$^\dagger$, Oscar Mickelin$^\ast$, and Amit Singer$^{\ast \S}$ 
		\thanks{T.B. is partially
			supported by the NSF-BSF award 2019752. {A.S. is partly supported by  AFOSR Award FA9550-
			20-1-0266, the Simons Foundation Math+X Investigator Award,  NSF BIGDATA Award
			IIS1837992, NSF Award DMS-2009753, and NIH/NIGMS Award R01GM136780-
			01}.}}
	\address{$^\dagger$School of Electrical Engineering, Tel Aviv University, Tel Aviv, Israel\\
		$^*$PACM, Princeton University, Princeton, NJ, USA \\
		{$^{\S}$ Department of Mathematics, Princeton University, Princeton, NJ, USA}}

	\ninept
	
\setlength{\abovedisplayskip}{4pt}
\setlength{\belowdisplayskip}{4pt}

	\maketitle
	\begin{abstract} 
		Motivated by the problem of determining the atomic structure of macromolecules using single-particle
		cryo-electron microscopy (cryo-EM), {we study the sample and computational complexities of the sparse multi-reference alignment (MRA) model: the
		problem of estimating a sparse signal from its noisy, circularly shifted copies.
	 Based on its tight connection to the crystallographic phase retrieval problem, we establish that if the
		number of observations is proportional to the square of the variance of the noise, then the sparse MRA problem is
		statistically feasible for sufficiently sparse signals. To investigate its computational hardness, we consider three types of computational frameworks: projection-based algorithms, bispectrum inversion, and convex relaxations. 
		We show that a state-of-the-art projection-based algorithm achieves the optimal estimation rate,  but its computational complexity is exponential  in  the  sparsity  level.   The  bispectrum  framework  provides a  statistical-computational  trade-off:  it  requires  more  observations (so its estimation rate is suboptimal), but  its  computational  load  is  provably  polynomial  in  the  signal’s length.  The convex relaxation approach provides polynomial-time algorithms (with a large exponent) that recover  sufficiently  sparse  signals  at  the  optimal estimation  rate.
		We conclude the paper by discussing potential statistical and algorithmic implications for cryo-EM.
		}
		
		\end{abstract}
	%
%

\begin{keywords}
	multi-reference alignment, sparse recovery, crystallographic phase retrieval, cryo-EM, expectation-maximization
\end{keywords}

	\section{Introduction}
	\label{sec:intro}
	We study the multi-reference alignment (MRA) problem of estimating a signal $x\in\R^L$
	from its $n$ circularly shifted, noisy copies
	\begin{equation} \label{eq:mra}
		y_i = R_{s_i}x+\varepsilon_i, \quad i=1,\ldots,n,
	\end{equation}
	where $R_s$ is the circular shift operator $(R_{s_i}x)[\ell]=x[(\ell-s_i)\bmod L]$, and the random variable $s$ is   drawn from a uniform distribution. The noise terms $\varepsilon_i$ are drawn i.i.d. from a normal distribution $\mathcal{N}(0,\sigma^2I)$.
	Given the observations $y_1,\ldots,y_n$, we can hope to estimate the signal only up to a global circular shift. 
	Thus,  the  goal is to estimate the orbit of the signal: the signal and all its circularly shifted copies $\{R_sx\}_{s=0}^{L-1}$. 
	
The MRA model is mainly motivated by the 3-D reconstruction problem in cryo-electron microscopy~\cite{bendory2020single}; see  Section~\ref{sec:cryoEM}.
In recent years, it was demonstrated that modeling the 3-D {molecular} structure  
as  a sparse mixture of Gaussians is highly effective for a variety of computational tasks, ranging from ab initio modeling to conformational variability analysis~\cite{chen2021deep,kawabata2018gaussian,joubert2015bayesian,rosenbaum2021inferring,zhong2021exploring}.
In particular, shrinking the width of the Gaussians to the scale of a single angstrom, one can interpret  each Gaussian  as representing a single atom, {and wider Gaussians provide coarser representations of features of the molecular structure}. 
This point of view, also termed the ``bag of atoms'' model, is also useful for analyzing the statistical properties of a  3-D structure consisting of {randomly positioned }atoms~\cite{wilson,Singer2021wilson}.
This sparse representation of the 3-D structure motivates this study of the sample and computational complexities of the sparse MRA model.

We model the target signal as an ensemble of shifted copies of an ``atom'' function $g$: 
\begin{equation} \label{eq:signal_model}
	x[n] = \sum_{i=1}^M g[n-m_i], 
\end{equation}
where  $m_i,\ldots,m_M$ is the series of unknown positions. 
{If $g$ is a Gaussian, then $x$ is a Gaussian mixture.}
Taking the discrete Fourier transform (DFT), we get 
\begin{equation}
	\hat{x}[k] = \hat{g}[k] \sum_{i=1}^M e^{2\pi\iota k m_i/L}, 
\end{equation}
where $\iota=\sqrt{-1}$, and $\hat{x}$ is the DFT of $x$. 
Assuming the DFT of~$g$, $\hat{g}[k]$, is {known and }bounded away from zero, we treat the signal simply as an $M$-sparse, binary signal, which is equivalent to considering~$g$ as a delta function. 

The MRA model behaves differently in low and high noise levels. 
If the noise level is low, one can estimate the circular shifts and then average the aligned observations~\cite{singer2011angular}.
However, estimating  the shifts accurately in the high noise regime is challenging, and thus one needs to bypass shift estimation, and estimate the signal directly. 
The high noise regime is of particular interest since
the signal-to-noise ratio in a typical cryo-EM dataset can be as low as 1/100~\cite{bendory2020single}. Thus, standard algorithms in the field are  based on methods that circumvent rotation estimation, such as the method of moments and expectation-maximization~\cite{scheres2012relion}.
	
This {paper} is focused on sparse MRA in the high noise regime $n,\sigma\to\infty$. 
In this regime, it was shown that the sample complexity of the MRA problem is controlled by the statistical moments of the observations (for finite $L$), e.g.,~\cite{perry2019sample,abbe2018estimation}.  Namely, assuming the  $d$-th order moment is the lowest-order moment of the observations that determines the orbit of the signal uniquely, then $n=\omega(\sigma^{2d})$ is a necessary condition for accurate estimation. This should be contrasted with the typical sample complexity in many signal processing tasks (as well as in MRA in the low noise regime), where $n=\omega (\sigma^{2})$ suffices. 
In Section~\ref{sec:sample_complexity} we discuss the ramifications of this result on the sample complexity of sparse MRA. Section~\ref{sec:computational_hardness}  studies the  computational complexity of the sparse MRA problem, and is the main contribution of this paper. 
In particular, we investigate three types of algorithms: projection-based, bispectrum inversion, and convex relaxations. 
We demonstrate that a state-of-the-are  projection-based algorithm achieves the optimal estimation rate, but its computational complexity is exponential in the sparsity level. 
The bispectrum framework provides a statistical-computational trade-off: it requires more observations (and thus its estimation rate is suboptimal), but its computational load is provably polynomial in the signal's length.
The convex relaxation approach provides polynomial-time algorithms that recover sufficiently sparse signals at the optimal estimation rate. However, the computational complexity of these algorithms scales with a large exponent,
and thus their applicability to  high-dimensional data, with the current implementations, is questionable.
We conclude the paper by  discussing potential statistical and algorithmic implications for cryo-EM. 
The code to reproduce all numerical experiments  appearing in this manuscript is publicly available at: \newline \url{https://github.com/TamirBendory/sparseMRA}

	\section{The sample complexity of sparse MRA}
	\label{sec:sample_complexity}
    
	As mentioned earlier, previous papers established that the sample complexity of the MRA problem in the high noise regime is determined by the statistical moments of the  observations.
	Under model~\eqref{eq:mra}, the moments are equivalent to invariants: features of the signal that remain unchanged under circular shifts. 
	For example, the mean of the signal is invariant under circular shifts. However, the mean is not enough to distinguish  different signal orbits,  and therefore we need to consider higher-order invariants.
	For non-sparse signals,	the same is true for the second-order invariant---the power spectrum (the Fourier transform of the autocorrelation function).
	Notably, it was shown that the third-order invariant, called the bispectrum, determines the orbit of a generic signal uniquely.
	This implies that  a necessary condition to accurately estimate a generic signal in the high noise regime is	$n=\omega(\sigma^6)$ observations. 
	
	While a non-sparse signal cannot be recovered from its power spectrum, it is not necessarily true for sparse signals---the focal point of this paper. In particular, if a signal can be recovered from its power spectrum, it implies a sample complexity of $n=\omega(\sigma^4)$, much lower than the sample complexity of non-sparse signals.
	Therefore, the question of sample complexity boils down to whether a sparse signal can be recovered uniquely  from its power spectrum. We note that in this case, the orbit of a signal is bigger and includes a reflection of the signal. 
	
	
	The problem of recovering a sparse signal from its power spectrum is of crucial importance for the X-ray crystallography technology, where the acquired data is equivalent to the power spectrum of a molecular structure of interest, and the goal is to recover the sparsely spread atom positions~\cite{elser2018benchmark}. 	
	Due to its practical importance, many algorithms for recovering a sparse signal from its power spectrum were designed (see Section~\ref{sec:computational_hardness}). 
	Nevertheless, only a few theoretical results are available. 
	In~\cite{ranieri2013phase}, the problem of recovering a signal from its power spectrum was first linked to the  classical  beltway and turnpike problems in combinatorics. 
	In particular, it was shown that when the support of the sparse signal is collision-free---namely,  the pairwise distances between support elements are distinct---then the orbit of the signal can be almost surely recovered uniquely. 
	The sparsity level of a typical collision-free set is $M=O(L^{1/3})$~\cite{ghosh2021multi}. 
	This result was greatly extended in~\cite{bendory2020toward}, which applied algebraic geometry tools  and established a conjecture that  the orbit of a signal is determined uniquely from its power spectrum, even for supports with collisions, as long as $M<L/2$.
	This result was rigorously proven in a limited regime of parameters, and conjectured to hold true for any ($M,L$). 
		Recently, 
	it was proven that there is a unique mapping for $M=O(L/\log^5(L))$ for symmetrical signals for large enough~$L$~\cite{ghosh2021multi}.
		For binary signals, the problem was introduced in~\cite{elser2017complexity}, and further investigated in \cite{bendory2020toward}. 
	Notably,  whether the power spectrum identifies uniquely a binary signal depends solely on the signal's difference set.
	
	\section{Computational hardness}
	\label{sec:computational_hardness}

In the previous section, we have established that in a certain sparsity regime, a signal can be recovered from its power spectrum, and therefore $n=\omega(\sigma^4)$ observations contain enough information to estimate a signal in the high noise regime. 
In this section, we focus on the computational question: can we \emph{efficiently} recover a sparse signal from $n=\omega(\sigma^4)$ observations?

{We explore this question using three types of algorithms. 
First, we implement a state-of-the-art projection-based algorithm. This algorithm uses the optimal number of observations, $n=\omega(\sigma^4)$, but its computational complexity scales exponentially fast with the sparsity level. 
Second, we consider an invariant-based algorithm that provides a computational-statistical trade-off:  a provable polynomial-time algorithm at the cost of suboptimal  $n=\omega(\sigma^6)$ number of observations. We also show that expectation-maximization-- the method of choice for MRA and cryo-EM~\cite{scheres2012relion}---provides the same trade-off. 
Finally, we consider polynomial-time, convex relaxation algorithms that provide accurate estimates for sufficiently sparse signals with only  $n=\omega(\sigma^4)$ observations. We note that the computational complexity of these algorithms scales with a large exponent, meaning that the algorithms in their current form cannot be {efficiently} applied to high-dimensional data. Thus, efficient implementations is an enticing future research direction. }


\subsection{Projection-based algorithms}
Motivated by X-ray crystallography,
the algorithmic aspects of recovering a sparse signal from its power spectrum were studied in a series of works,  see for example~\cite{fienup1982phase,elser2003phase,elser2007searching,elser2018benchmark}. 
These algorithms are based on two easy-to-compute projection operators. 
Importantly, these algorithms are not of the alternating-projections type that tends {to stagnate}, but use the projections in a more intricate way, sharing similarities with  the Douglas-Rachford and { the alternating direction method of multipliers} frameworks~\cite{lindstrom2021survey}.

To be more concrete, we consider here the relaxed-reflect-reflect (RRR) algorithm~\cite{elser2017complexity,elser2018benchmark}.  
RRR builds on two projection operators: the first encodes  prior information on the signal, while the second is data-dependent.  
For the problem at hand,  the first projection~$P_1$ enforces sparsity (namely, keeps only the $M$ largest  entries of the signal), while the second projection~$P_2$ imposes the measured power spectrum (this projection can be computed in $O(L\log L)$ operations).
The RRR iterations read 
\begin{equation}
	x\mapsto x + \beta (P_2(2P_1(x)-x) - P_1(x)),
\end{equation}
where $\beta\in(0,2)$ is a parameter. 
For $\beta=1$, RRR coincides with Douglas-Rachford. 

\begin{figure}
	\centering
	\includegraphics[width=0.7\columnwidth]{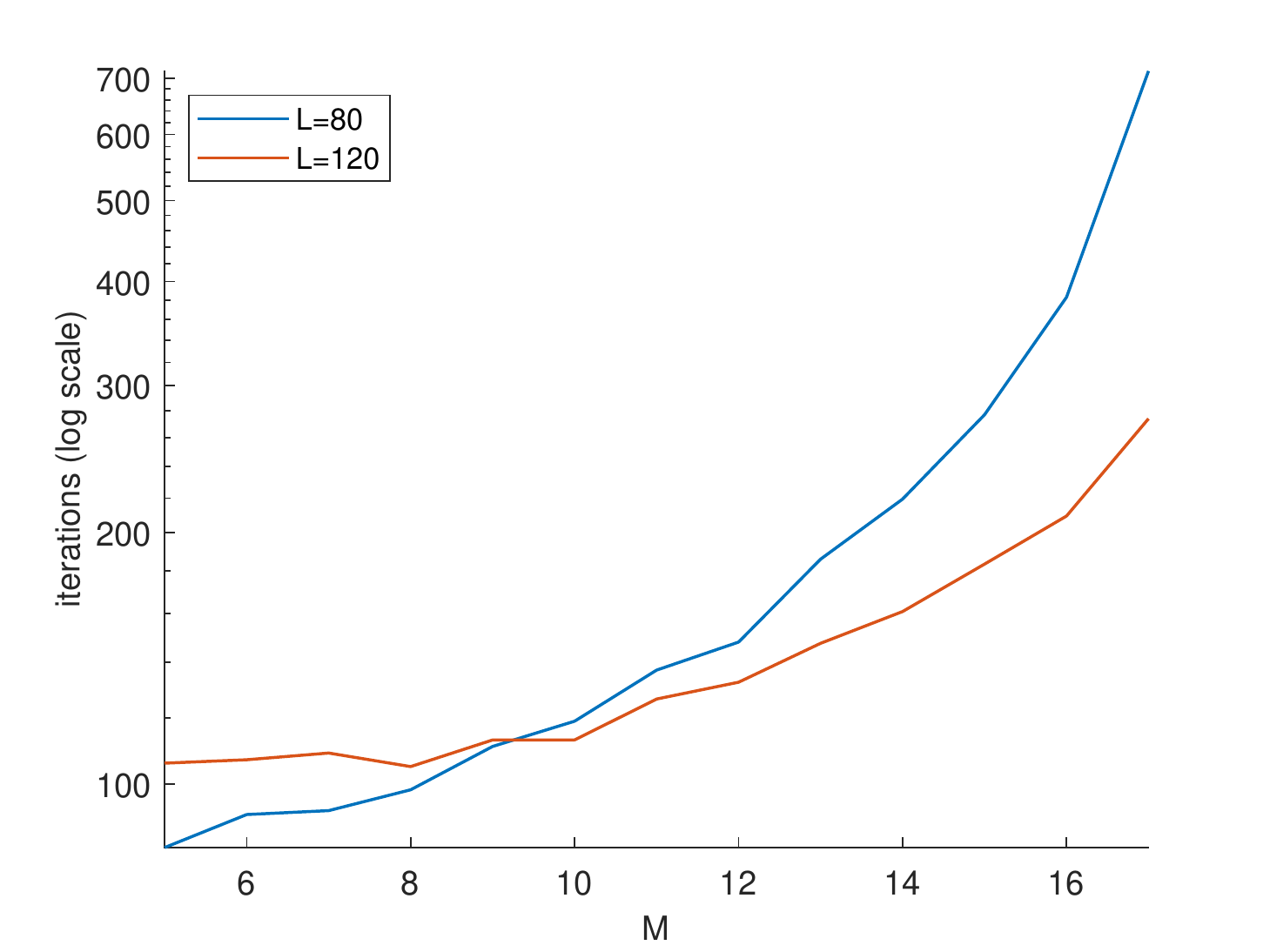}
	\caption{\label{fig:rrr} The median number of iterations required for {RRR} to find an $M$ sparse signal of length $L$ from its power spectrum. Evidently, the number of iterations grows exponentially fast with the sparsity level, indicating that the computational complexity of the algorithm (and perhaps of the problem itself) is exponential in the sparsity level.} 
\end{figure}

A crucial property of RRR is that in the absence of noise, it halts only when it finds a sparse signal with the measured power spectrum. 
Figure~\ref{fig:rrr} presents the number of iterations required for RRR with $\beta=1/2$ to find a solution as a function of the sparsity level~$M$. 
In the experiment, we generated a binary, sparse signal of length $L=80,120$  with $M$ non-zero entries. $M$ is assumed to be known since it can be read off from the first moment of the signal, or, equivalently, from the zero frequency of the power spectrum. For each pair $(L,M)$, we ran 500 trials, each with a fresh signal, and counted iterations till convergence. We stopped the algorithm when the relative error between the measured power spectrum and the power spectrum of the estimated signal, after the sparsity projection, dropped below $10^{-5}$.



 
Evidently, the number of RRR iterations grows  exponentially fast in the sparsity level, agreeing with previous  results in the field~\cite{elser2018benchmark,elser2017complexity}.
Since RRR provides the best-known results for such problems (at least for high-dimensional data), these results indicate  that perhaps recovering a sparse signal from its power spectrum is a hard problem: finding a solution  requires an exponential number of operations.

\subsection{Bispectrum inversion and expectation-maximization}

\noindent\textbf{Bispectrum inversion.}
The bispectrum of a signal $x\in\C^L$ consists of all triple products $\hat{x}[k_1]\hat{x}[k_2]\hat{x}[-k_1-k_2]$ for $k_1,k_2=0,\ldots,L-1$, where $\hat{x}$ is the DFT of $x$. By construction, the bispectrum is invariant under circular shifts: the bispectra of $x$ and $R_sx$ are the same, for any $s\in\mathbb{Z}$. 
The bispectrum inversion method is a special case of the method of moments, and consists of two stages. 
 First, we compute the bispectra of all observations and average them to obtain an estimate of the bispectrum of the signal. Since the bispectrum is a third-order invariant, its variance is proportional to $\sigma^6$, and therefore $n=\omega(\sigma^6)$ observations are required for an accurate estimate; this coincides with the sample complexity for generic (non-sparse) signals in the high noise regime.
 The second stage consists of  estimating a signal from its bispectrum. While the problem is non-convex, there exist several efficient techniques to recover a signal from its bispectrum (some with theoretical guarantees)~\cite{bendory2017bispectrum,perry2019sample}. 
 
\noindent\textbf{Expectation-maximization.} Expectation-maximization (EM)
  is a popular algorithm to maximize the posterior distribution in MRA and cryo-EM. For the MRA model~\eqref{eq:mra}, the log-posterior distribution reads:
	\begin{equation}
		\log p(x|y_1,\ldots,y_n) \! = \hspace{-3pt} \sum_{i=1}^n \hspace{-1pt} \log \hspace{-2.5pt} \sum_{s=0}^{L-1} \hspace{-1pt} \mathcal{N}(y_i \hspace{-0.75pt} - \hspace{-0.75pt} R_sx, \sigma^2 I) + \log p(x),
	\end{equation}
	where $p(x)$ is the prior on $x$. 
	We consider a  Bernoulli prior with parameter $q$ so that
	\begin{equation}
		p(x) = \prod_{i=0}^{L-1}p(x[i])=\prod_{i=0}^{L-1}q^{x[i]}(1-q)^{1-x[i]}=q^M(1-q)^{L-M},
	\end{equation}
	where $M = \sum_{i=0}^{L-1}x[i]$. 
	Then, the EM iterations read
	\begin{equation}
		x_{t+1} = \frac{1}{n}\sum_{i=1}^n\sum_{s=0}^{L-1}w_{i,s}R_{s}^{-1}y_i + \frac{2\sigma^2}{n}\log\frac{q}{1-q},
	\end{equation}
    where the weights $w_{i,s}$ are given by 
	\begin{equation}
		w_{i,s} = \frac{e^{-\frac{1}{2\sigma^2}\|y_i-R_sx_t\|^2}}{\sum_{s=0}^{L-1}e^{-\frac{1}{2\sigma^2}\|y_i-R_sx_t\|^2}}.
	\end{equation}
	Note  that the contribution of the prior is negligible since $\sigma^2\ll n$.
	
\begin{figure}
			\centering
		\includegraphics[width=.75\columnwidth]{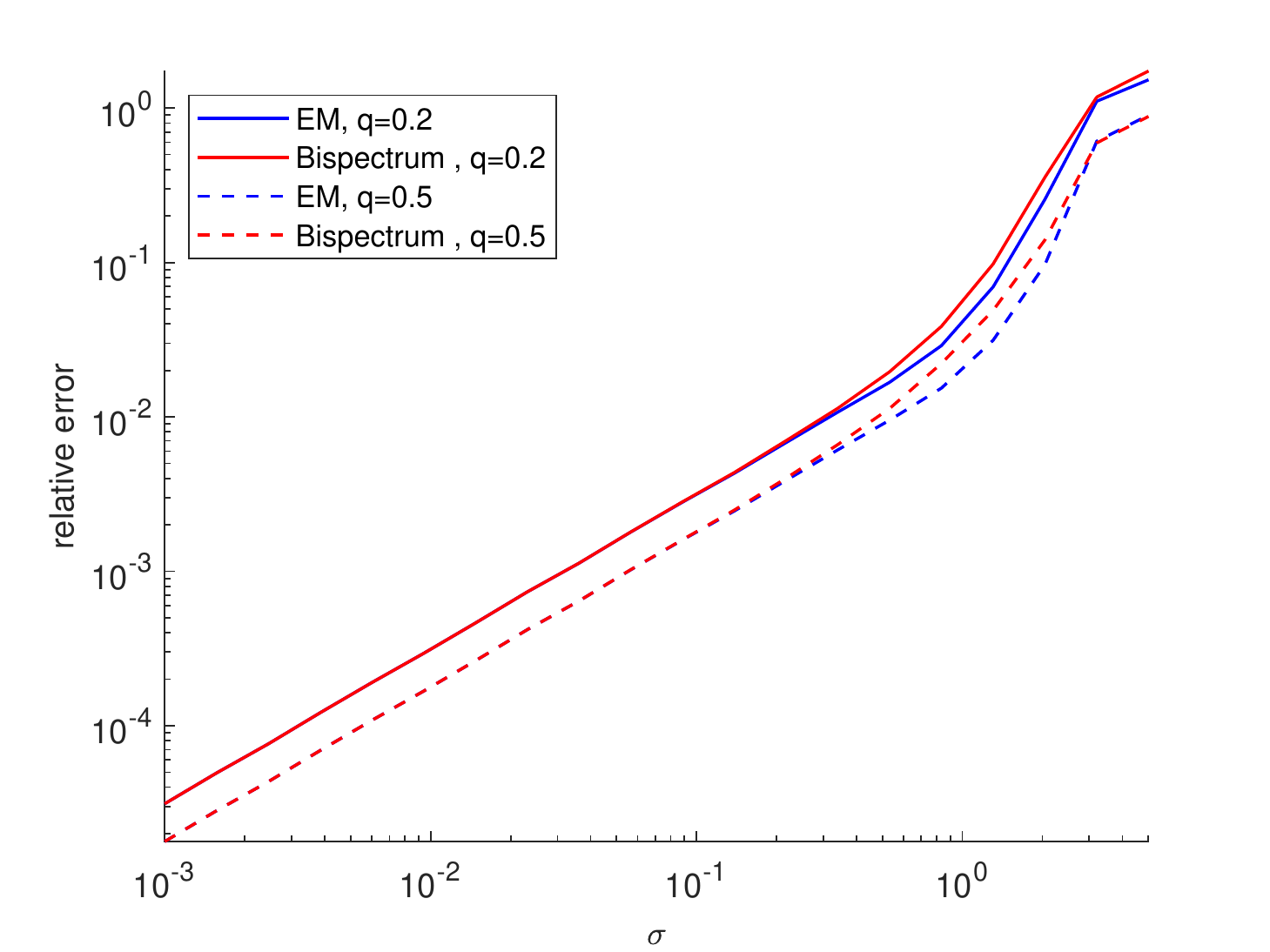}
		
		\caption{\label{fig:em_bis} Relative  error of estimating a  signal of length $L=60$ with {entries} drawn from a Bernoulli distribution with parameter $q$. We used  the bispectrum inversion and EM algorithms to recover the signal from $n=5000$ MRA observations~\eqref{eq:mra}.
		The fact that the error curves of EM follow the same trend as the bispectrum error suggests  that the EM---the standard algorithm for MRA and cryo-EM---does not achieve the optimal estimation rate for sparse models.  	
	}
	\end{figure}

	\noindent \textbf{Numerical experiments.}
	{Figure~\ref{fig:em_bis}} shows the average recovery error (over 100 trials) as a function of the noise level  for EM and bispectrum inversion. 
	Ground-truth signals of length $L=60$ were drawn from a Bernoulli distribution with parameters $q=0.2,0.5$. 
	Then, we generated $n=5000$ observations according to~\eqref{eq:mra}. The bispectrum inversion algorithm was solved using a non-convex least-squares algorithm proposed in~\cite{bendory2017bispectrum,boumal2018heterogeneous}. The EM iterations were halted when the difference between consecutive iterations dropped below $10^{-7}$. 
	In the low noise regime, the slope of all error curves is $1$, which is the estimation rate of averaging Gaussian i.i.d.\ variables. However, around $\sigma\approx 1$, the slope of the error curves  increases dramatically to around 3.
	For the bispectrum, this phenomenon is easily explained: the bispectrum consists of triple products of noisy terms, and thus its standard deviation in the high noise regime is proportional to $\sigma^3$. 
	However, interestingly, also the error curve of EM---which does not build explicitly upon the moments, but aims to maximize the posterior distribution directly---follows the same trend. 
	
	The results of 	{Figure~\ref{fig:em_bis}} suggest that EM, which is the standard  algorithm for MRA and cryo-EM, does not achieve the optimal estimation rate for sparse signals. 
	{In particular, they imply that perhaps by using phase retrieval algorithms one can recover signals with fewer observations than EM, at the cost of exponential running time. 
	In addition, the results may indicate a \emph{computational-statistical gap}: no algorithm can be both statistically and computationally optimal (a similar phenomenon was observed in the related problem of  heterogeneous MRA~\cite{boumal2018heterogeneous,bandeira2017estimation,bandeira2018notes}.) 
	In the next section, we devise polynomial-time algorithms that recover sufficiently sparse signals with merely $n=\omega(\sigma^4)$, suggesting that if the gap indeed exists, it holds only for $M\geq M_0$ for some $M_0$, namely, only for dense signals.
	}	
	
\subsection{Convex relaxations for binary phase retrieval} 
We consider a convex relaxation technique based on solving a semidefinite program (SDP) in a higher-dimensional space.
This strategy is popular in the phase retrieval literature~\cite{candes2015phase,jaganathan2017sparse}, as well as in cryo-EM~\cite{singer2011three}. 

Let $y=|Fx|^2,$ where $x\in\C^L$ is the target signal, $F\in\C^{L\times L}$ is the DFT matrix, and the absolute value should be considered entry-wise. 
Let $X:=xx^*$ be a Hermitian, rank 1, positive semidefinite matrix, where $x^*$ is the conjugate-transpose of $x$. 
 Let $f_i$ be the $i$-th column of $F$, and  $F_i:=f_if_i^*$. Equipped with these definitions, we can write $\tr(F_iX)=y[i]$.
Since the signal is binary, we also have $x[i]=x^2[i]$. 
Assuming $x[0]=1$, this is equivalent to requiring $X[0,0]=1$, and $X[i,i]=X[0,i]$ for  $i=0,\ldots,L-1$.
Therefore, we write the binary phase retrieval problem as 
\begin{equation}
	\begin{split}
		\text{find}\quad  {X\in \mathcal{S}^{L\times L}} \quad \text{subject to} \quad X\in\Omega,\,
								\text{rank}(X)=1,\,
	\end{split}
\end{equation}
where $\mathcal{S}^{L\times L}$ is the space of symmetric matrices of size $L\times L$, and~$\Omega$ is the set of signals that satisfy the convex constraints $X\succeq 0$, $\tr(F_iX) = y[i],   X[i,i]=X[0,i],   X[0,0]=1,   X[i,j]\geq 0,$ for $i,j=0,\ldots,L-1$. 
In order to convexify the problem, we propose to omit the rank constraint and solve the following SDP relaxation
\begin{equation} \label{eq:sdp}
	\begin{split}
		\min_{X\in \mathcal{S}^{L\times L}}\tr(RX) \quad \text{subject to} \quad X\in\Omega, 
	\end{split}
\end{equation}
where $R\in\R^{L\times L}$ is a random matrix 
that, with probability 1, favors one of the equivalent solutions; otherwise, the convex program may converge to any convex combination of  solutions.

\begin{figure}
	\centering
	\includegraphics[width=1\columnwidth]{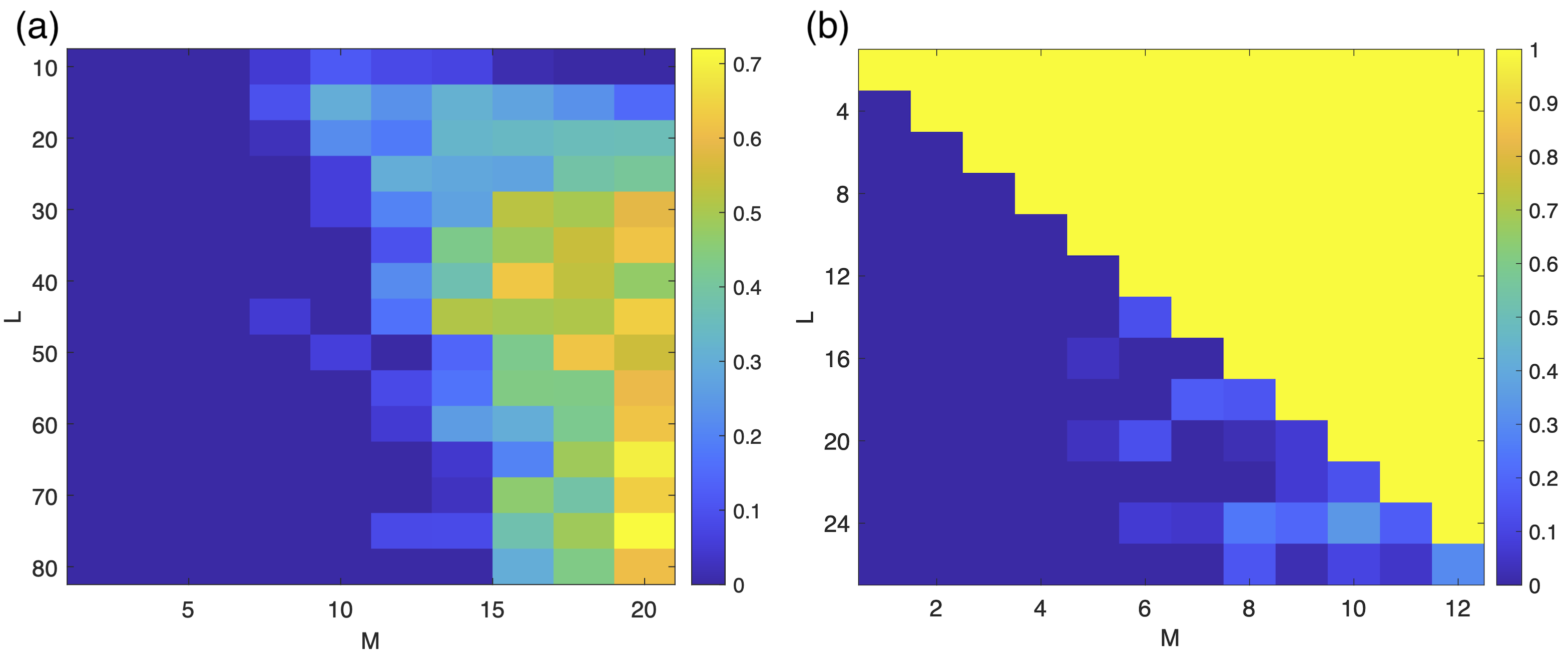}
	\caption{\label{fig:pr} Average recovery error of (a) the SDP in~\eqref{eq:sdp} and (b) the SoS implementation, as a function of the signal length $L$ and sparsity level $M$. Up to some sparsity level, both convex programs estimate the signal accurately. 
	}
\end{figure}

{Figure~\ref{fig:pr}} shows the average recovery error (over 10 trials) for recovering a binary signal from its power spectrum using the convex relaxation~\eqref{eq:sdp} as a function of $(L,M)$. The SDP was solved using CVX~\cite{cvx}. 
The results suggest that, up to some sparsity level, it might be possible to solve the binary phase retrieval problem in polynomial time. 
However, the running time of the algorithm scales badly with~$L$, rendering it impractical for experimental datasets composed of  thousands of elements.
Considering efficient solvers, such as the Burer-Monteiro approach~\cite{boumal2016non}, might mitigate the computational burden and is an interesting future research direction. 
	

We  also implemented a  two levels sum-of-squares (SoS) hierarchy~{\cite{henrion2009gloptipoly,cosse2020stable}}. {SoS-type algorithms reduce polynomial optimization problems to SDPs by introducing new variables for each monomial appearing in the cost function. The techniques impose positive semidefiniteness constraints on the moments of these variables, up to degree twice the level of the hierarchy, and global optimality is ensured, provided the level is chosen sufficiently large. The size of the optimization problem grows polynomially in the level, but with a large exponent.}
The results of the SoS were  better than the SDP, but the running time of the algorithm is even more prohibitive  (although polynomial). The average recovery error (over 10 trials) as a function of $(L,M)$ appears in Figure~\ref{fig:pr}.
Recently, there were intriguing attempts to replace SoS with efficient spectral methods that enjoy similar properties, see for example~\cite{hopkins2016fast}.
It is an interesting future research direction to explore if similar ideas can be applied  to the problem of recovering a sparse signal from its power spectrum.

We also mention that binary MRA can additionally be approached as an instance of a binary matrix decomposition problem, since the second moment of the signal in~\eqref{eq:mra} can be written as a sum of outer products of binary vectors. Recovering the binary decomposition of the moment is equivalent to recovering the orbit of the signal. Recently, there has been an interest in SDP-based approaches for binary matrix decompositions \cite{kueng2021binary}, and yet another interesting avenue for future research consists in exploring adaptations of these methods to sparse MRA.

	\section{Potential implications to cryo-EM} \label{sec:cryoEM}
	
	Under some simplifying assumptions, it was recently shown that the sample complexity of the cryo-EM problem in the high noise regime is $n=\omega(\sigma^6)$~\cite{bandeira2017estimation}. However, similarly to the analysis of this paper, representing the structure as a ``bag of atoms'' or as a Gaussian mixture model alters the statistical model, and thus the sample complexity will naturally depend on the sparsity level of the model.

	In particular, the second moment (the analog of the power spectrum in MRA) of cryo-EM projections was first calculated  by 
	Zvi Kam under the assumption of a uniform distribution of 3-D rotations~\cite{kam1980reconstruction}. In particular, Kam showed  that the second moment determines the structure up to a set of orthogonal matrices composed of the spherical harmonic coefficients of the sought molecular structure. 
	These missing orthogonal matrices can be thought of as the analog of the missing Fourier phases in the crystallographic phase retrieval problem. 
	Several methods were suggested to recover the missing orthogonal matrices, such as using a homologous structure~\cite{bhamre2015orthogonal}, or part of the raw data~\cite{levin20183d}. 
	However, we postulate that a sparse representation of the 3-D structure (under some basis) will allow determining the missing orthogonal matrices uniquely. 
	In particular, it might be possible to  design a projection-based algorithm, similarly to RRR, for cryo-EM. In this case, one projection will enforce sparsity by keeping only the leading $M$ entries of the signal in the space domain, while the second operator will project the signal estimate (in the spherical harmonics domain) onto the space of orthogonal matrices; this projection can be efficiently implemented using singular-value decomposition (as the solution of the Procrustes problem). 
	
	The potential of using RRR in cryo-EM is enormous. 
	The standard algorithms in cryo-EM are based on EM. However, by drawing analogy from MRA, it seems that they need $n=\omega(\sigma^6)$ observations to achieve an accurate estimate.  The second moment can be estimated with only $n=\omega(\sigma^4)$ observations, and thus RRR may be able to recover structures with  fewer observations; this would come at the cost of exponential running time. 
	While all these statements are conjectural by nature, we aim to verify the huge potential of applying phase retrieval algorithms to cryo-EM datasets in future research. 
	\newpage
	\bibliographystyle{IEEEbib}

\end{document}